\documentclass[12pt,reqno]{amsart}
\usepackage[utf8]{inputenc}
\usepackage[T2A]{fontenc}
\usepackage{xcolor}
\usepackage{amsmath,amsthm,amscd,amsfonts,amssymb,graphicx,color}
\usepackage[bookmarksnumbered,colorlinks,plainpages]{hyperref}
\usepackage[english,ukrainian]{babel}
\usepackage{calc}
\usepackage{fancybox}
\usepackage{geometry}
\usepackage{array,tabularx}
\usepackage{fancyhdr}
\hypersetup{colorlinks=true,linkcolor=blue,urlcolor=blue}

\theoremstyle{definition}

\theoremstyle{remark}

\begin{document}
\setcounter{page}{1}
\fancyhead[RO]{\centering\colorbox{white}{Siasiev A. Coupled Thermomechanical Model with Volterra Memory}}
\fancyhead[LE]{\centering\colorbox{white}{{\fontencoding{T2A}\selectfont Сясєв А. Зв’язана термомеханічна модель з пам’яттю типу Вольтерри}}}
\fancyfoot[C]{\raisebox{0.5\height-\baselineskip}{\thepage}}

\begin{flushright}
\ovalbox{\bf М І Ф}
\end{flushright}
\noindent \scalebox{0.95}[1]{\bf Математика, Інформатика, Фізика: Наука та Освіта, Том 3, № 2 (2026), с. 1--??.}\hfill \\
\scalebox{0.95}[1]{\it Mathematics, Informatics, Physics: Science and Education, Volume 3, No. 2 (2026), pp. 1--??.}\hfill \\
{Journal homepage: \url{https://vspu.net/miph/index.php/journal}}\hfill
\noindent\rule{\linewidth}{0.2pt}

\centerline{}
UDC 517.968.7
\vspace{2mm}
\begin{center}
{\bf\Large A Coupled Thermomechanical Model with Volterra Memory: Influence of the Kernel Class on Relaxation Asymptotics}\par
\vspace{4mm}
\noindent{\bf Andrii Siasiev}\\
\vspace{2mm}
\noindent{\scriptsize Oles Honchar Dnipro National University, Department of Mathematical Analysis and Optimization, Dnipro, Ukraine}\\
\noindent{\scriptsize syasevandr@gmail.com}\\
\noindent{\scriptsize \url{https://orcid.org/0000-0002-0654-7360}}
\end{center}
\vspace{2mm}
\noindent\rule{\linewidth}{0.2pt}
\author[]{}
\title[]{}
\renewcommand{\datename}{}
\date{e-ISSN 3041-1955\quad \hspace{30mm} DOI: [assigned by the journal] 
© 2026 Author, Creative Commons Attribution 4.0 International Licence}

\renewcommand{\abstractname}{\quad\it Abstract}
\begin{abstract}
The aim of this study is to determine how the class and asymptotic behaviour of a Volterra memory kernel influence relaxation in a coupled thermomechanical model. The model describes the mutual interaction between an internal mechanical variable and temperature through temperature-dependent material properties and mechanical dissipation. The analysis combines a Volterra integral formulation, a spectral representation of the memory mechanism, thermodynamic decomposition of the hereditary power, local well-posedness analysis, linearization near equilibrium, and Laplace-transform asymptotic arguments. The spectral representation separates the energy stored in the relaxation modes from irreversible hereditary dissipation and establishes the thermodynamic consistency of the memory contribution. Under minimal assumptions on the constitutive functions and the kernel, local existence and uniqueness of solutions are obtained; for thermodynamically dissipative memory and a non-negative external heat source, positivity of the absolute temperature is preserved. The main result concerns the linearized dynamics near equilibrium: with the thermomechanical coupling kept unchanged, an exponential kernel produces exponential relaxation, whereas a power-law kernel with a long tail produces algebraic relaxation. A numerical comparison performed for the same set of dimensionless thermomechanical parameters illustrates these two asymptotic regimes. The results show that changing the class of the memory kernel alone can change the qualitative long-time relaxation law of the coupled system.

\end{abstract} \maketitle
\noindent\rule{\linewidth}{0.2pt}

\begingroup
    \setlength{\parskip}{0pt}
    \section{\large\bf 1. Introduction}
	
	Coupled thermomechanical processes arise in a wide range of problems involving deformable solids, where mechanical deformation affects the thermal state and temperature variations, in turn, modify the mechanical response. Such mutual interaction may become particularly significant in nonstationary processes accompanied by relaxation, crystallization, or structural evolution. Mathematical models accounting for the mutual influence of temperature and mechanical fields have therefore received considerable attention; in particular, coupled formulations have been employed to describe the stress--strain state of materials under thermally driven evolution \cite{siasiev-et-al-2020}. Modern thermoelastic models extend this framework by incorporating various relaxation mechanisms and non-Fourier effects \cite{bazarra-fernandez-quintanilla-2023}.

In many materials, however, the current response is determined not only by the instantaneous state but also by its previous evolution. Such hereditary effects are naturally represented by convolution-type or Volterra integral operators. Memory terms have consequently become an important component of thermoelastic and viscoelastic models. Recent studies have established well-posedness and stability properties for thermoelastic systems with infinite memory and delay \cite{hao-zhang-2021}, thermoelastic Timoshenko systems involving memory and second-sound effects \cite{moumen-et-al-2021}, and more elaborate systems combining thermal coupling, diffusion, memory, and distributed delay \cite{laribi-et-al-2024}. These results demonstrate that memory can substantially affect the dissipative behaviour of coupled systems.

A particularly relevant question concerns the relation between the properties of the memory kernel and the long-time behaviour of solutions. General decay estimates have been obtained for viscoelastic systems with past history and general relaxation kernels \cite{al-mahdi-2020,al-mahdi-al-gharabli-guesmia-messaoudi-2021}. In the thermoelastic setting, the asymptotic decay rate has also been related explicitly to the behaviour of the relaxation function at infinity. This has been demonstrated for porous thermoelastic systems with infinite memory \cite{al-mahdi-al-gharabli-messaoudi-2021}, microbeam models governed by the Gurtin--Pipkin thermal law \cite{chen-liu-chen-2022}, and thermoelastic Timoshenko systems with infinite memory \cite{almutairi-mukiawa-2024}. More recently, a unified stability framework combining thermal coupling, second sound, and nonlinear infinite memory has been developed in \cite{boulaaras-guefaifia-2026}. Thus, the dependence of stability and decay estimates on the relaxation kernel is already an established feature of modern hereditary thermoelasticity.

At the same time, different kernel classes encode qualitatively different memory mechanisms. Exponentially decaying kernels correspond to fading memory with a characteristic relaxation scale, whereas slowly decaying kernels may retain the influence of remote history. Completely monotone kernels provide a broad mathematical framework encompassing mixtures of exponential relaxation modes and related Volterra structures \cite{alfonsi-kebaier-2024}. Power-law behaviour, in turn, is characteristic of long-memory responses and is closely connected with fractional viscoelastic models possessing a broad spectrum of relaxation time scales \cite{bonfanti-kaplan-charras-kabla-2020}. Consequently, the asymptotic tail of the kernel may be expected to play a decisive role in determining the relaxation regime.

The studies cited above primarily address stability and decay for a prescribed thermoelastic or viscoelastic system under suitable assumptions on its memory kernel. Comparatively less attention has been paid to a complementary question: how the long-time relaxation law changes when the class of the memory kernel itself is varied while the underlying thermomechanical coupling is kept unchanged. A minimal coupled formulation is particularly suitable for this purpose, since it allows the effect of the memory kernel to be separated from additional spatial, geometric, and dissipative mechanisms present in more elaborate continuum models.

The aim of the present paper is therefore to investigate a coupled thermomechanical model with Volterra memory and to determine how the class and asymptotic behaviour of the memory kernel influence the relaxation dynamics. The distinguishing feature of the approach is that different relaxation regimes are analysed within the same thermomechanical structure, with the kernel class acting as the principal variable factor. This makes it possible to compare short-memory and long-memory mechanisms on a common mathematical basis and to identify the corresponding long-time asymptotic behaviour without introducing unnecessary additional components into the model.

The paper is organized as follows. First, the coupled thermomechanical model with Volterra memory is formulated and the assumptions ensuring its mathematical consistency are specified. The basic analytical properties of the problem are then established. Particular attention is subsequently given to the asymptotic relaxation behaviour generated by different classes of memory kernels. Finally, the analytical conclusions are illustrated numerically and their thermomechanical interpretation is discussed.
\endgroup
	
	\section{\large\bf 2. Thermomechanical model with Volterra memory}
	
	We consider a homogeneous representative element of a thermoviscoelastic
material subjected to a uniaxial deformation process. Spatial heterogeneity
is neglected, and both the temperature and the internal mechanical state
are assumed to be uniform within the element. The resulting formulation
is therefore a lumped-parameter model depending only on time. The total
strain $\varepsilon(t)$ is prescribed as an external mechanical input.

Let $y(t)$ denote the internal viscous (relaxational) component of the
strain. The total strain is decomposed into elastic and internal parts,
\[
\varepsilon(t)=\varepsilon_e(t)+y(t),
\qquad
\varepsilon_e(t)=\varepsilon(t)-y(t).
\]
For small deformations, the elastic stress is taken in the form
\[
\sigma_e(t)=E(T(t))[\varepsilon(t)-y(t)],
\]
where $E(T)>0$ is the temperature-dependent elastic modulus. Thus, $y(t)$ has a direct mechanical interpretation as an internal
delayed or irreversible rearrangement of the material rather than as
an auxiliary mathematical variable. Such hereditary viscoelastic
effects are also relevant in deformation processes associated with
the gradual build-up of material \cite{siasiev-2001}.

The second state variable is the absolute temperature $T(t)$ of the
same material element. The variables $y(t)$ and $T(t)$ are coupled in
both directions: the temperature affects the elastic and viscous
material characteristics, whereas part of the mechanical energy
associated with the evolution of the internal strain is converted
into heat.

The elastic stress is assumed to provide the driving force for the
evolution of the internal strain. Its action is opposed by instantaneous
viscous resistance, an internal restoring or hardening resistance,
and a hereditary resistance associated with the previous relaxation
history. Accordingly, the mechanical balance is written as
\begin{equation}
\eta(T(t))\dot y(t)+R(y(t),T(t))+\sigma_K(t)
=
E(T(t))[\varepsilon(t)-y(t)],
\label{eq:mechanical-balance}
\end{equation}
where $\eta(T)>0$ is the temperature-dependent viscosity,
$R(y,T)$ is the internal restoring or hardening resistance, and
$\sigma_K$ denotes the hereditary stress. No particular polynomial
form of $R$ is imposed in the general formulation.

The hereditary stress appearing in \eqref{eq:mechanical-balance} is
represented through the history of the internal strain rate as
\begin{equation}
\sigma_K(t)=\int_0^t K(t-s)\dot y(s)\,ds,
\label{eq:hereditary-stress}
\end{equation}
where $K(t)$ is the relaxation kernel. The Volterra term
\eqref{eq:hereditary-stress} admits a natural interpretation in terms
of a spectrum of hidden relaxation mechanisms. Let $z_\tau(t)$ be an
internal variable associated with a characteristic relaxation time
$\tau>0$ and satisfying
\[
\tau\dot z_\tau(t)+z_\tau(t)=\dot y(t).
\]
For a zero initial state,
\[
z_\tau(t)=\frac{1}{\tau}\int_0^t
e^{-(t-s)/\tau}\dot y(s)\,ds.
\]
If the hereditary stress is generated by a spectrum of such relaxation
modes, then
\[
\sigma_K(t)
=
\int_0^\infty a(\tau)z_\tau(t)\,d\tau
=
\int_0^t K(t-s)\dot y(s)\,ds,
\]
with
\begin{equation}
K(t)=\int_0^\infty
\frac{a(\tau)}{\tau}e^{-t/\tau}\,d\tau.
\label{eq:spectral-kernel}
\end{equation}
Representation \eqref{eq:spectral-kernel} shows that the Volterra
memory can be interpreted as a macroscopic manifestation of a spectrum
of internal relaxation processes. It will also provide the basis for
the thermodynamic interpretation of the hereditary contribution in
Section~2.

Substitution of \eqref{eq:hereditary-stress} into
\eqref{eq:mechanical-balance} gives the basic nonlocal evolution
equation
\begin{equation}
\eta(T(t))\dot y(t)
+R(y(t),T(t))
+\int_0^t K(t-s)\dot y(s)\,ds
=
E(T(t))[\varepsilon(t)-y(t)].
\label{eq:nonlocal-mechanical}
\end{equation}
Possible nonlinearity of model \eqref{eq:nonlocal-mechanical} is
retained through the temperature dependence of $E$ and $\eta$ and
through the constitutive function $R(y,T)$, without prescribing an
additional nonlinear term solely for analytical convenience.

For the thermal part, we use the averaged energy balance
\[
C\dot T(t)
=
-h[T(t)-T_{\mathrm{env}}]
+q_{\mathrm{ext}}(t)
+D(t),
\]
where $C>0$ is the volumetric heat capacity, $h\geq0$ is the
heat-exchange coefficient, $T_{\mathrm{env}}>0$ is the ambient
temperature, $q_{\mathrm{ext}}(t)$ is an external volumetric heat
source, and $D(t)\geq0$ denotes the mechanical dissipation. The
instantaneous viscous contribution is
\[
D_\eta(t)=\eta(T(t))\dot y^{\,2}(t).
\]
Writing the hereditary contribution as $D_K[y](t)$, the thermal
equation takes the form
\begin{equation}
C\dot T(t)
=
-h[T(t)-T_{\mathrm{env}}]
+q_{\mathrm{ext}}(t)
+\eta(T(t))\dot y^{\,2}(t)
+D_K[y](t).
\label{eq:thermal-equation}
\end{equation}
The functional $D_K[y]$ in \eqref{eq:thermal-equation} is not
introduced phenomenologically. Its thermodynamic form will be derived
from the spectral representation \eqref{eq:spectral-kernel} in
Section~2, where the energy temporarily stored in the relaxation modes
will be separated from the irreversible hereditary dissipation.

The formulation is dimensionally consistent: every term in
\eqref{eq:nonlocal-mechanical} has the dimension of stress, whereas
every term in \eqref{eq:thermal-equation} has the dimension of power
per unit volume.

For the subsequent analysis, only the minimal assumptions required
for physical admissibility and mathematical consistency are imposed.
The temperature is restricted to the physical region $T>0$, with
\[
C>0,\qquad h\geq0,\qquad T_{\mathrm{env}}>0.
\]
The material functions satisfy
\[
E(T)>0,\qquad \eta(T)>0,
\]
and $E$ and $\eta$ are assumed to be continuous and locally Lipschitz
on the physically admissible temperature interval. The function
$R(y,T)$ is continuous and locally Lipschitz in its arguments and
satisfies
\[
R(y,T)y\geq0,
\]
so that the internal resistance does not amplify the internal
deformation.

For the basic Volterra analysis, the relaxation kernel is assumed
to satisfy
\[
K\in L^1_{\mathrm{loc}}(\mathbb{R}_+),\qquad
K(t)\geq0,\qquad K'(t)\leq0,
\]
where the derivative condition is understood whenever the corresponding
regularity is available. Stronger requirements, in particular complete
monotonicity, are not imposed at this stage. They will be introduced
only when required by the thermodynamic representation or the
asymptotic analysis.

The prescribed total strain $\varepsilon(t)$ is assumed to have the
regularity required by the evolution problem. The initial conditions
are
\begin{equation}
y(0)=y_0,\qquad T(0)=T_0>0.
\label{eq:initial-conditions}
\end{equation}
Conditions \eqref{eq:initial-conditions} specify the initial state of
the coupled problem. No separate prehistory for $t<0$ is introduced
because the Volterra integral in \eqref{eq:hereditary-stress} starts
at $t=0$.

These conditions constitute the basic set of assumptions used below.
No additional restrictions on the constitutive functions or on the
memory kernel are imposed unless they are explicitly required for a
particular analytical result.
	
	
	\section{Thermodynamic consistency and kernel classes}
	
	The hereditary term introduced in Section~1 represents a mechanical
memory mechanism and therefore requires a thermodynamic interpretation.
In particular, the contribution $D_K[y](t)$ in
\eqref{eq:thermal-equation} must be distinguished from the energy
temporarily stored in the internal relaxation modes. This distinction
can be made directly by using the spectral representation
\eqref{eq:spectral-kernel}.

For each relaxation time $\tau>0$, let $z_\tau(t)$ satisfy
\[
\tau\dot z_\tau(t)+z_\tau(t)=\dot y(t).
\]
Multiplication by $a(\tau)z_\tau(t)$ gives
\[
a(\tau)z_\tau(t)\dot y(t)
=
a(\tau)z_\tau^{\,2}(t)
+
a(\tau)\tau z_\tau(t)\dot z_\tau(t).
\]
Integrating over the relaxation spectrum and using the representation
of the hereditary stress yields
\begin{equation}
\sigma_K(t)\dot y(t)
=
\frac{d}{dt}\Psi_K(t)+D_K[y](t),
\label{eq:memory-power-decomposition}
\end{equation}
where
\[
\Psi_K(t)
=
\frac{1}{2}\int_0^\infty
a(\tau)\tau z_\tau^{\,2}(t)\,d\tau
\]
is the energy temporarily stored in the internal relaxation modes, and
\begin{equation}
D_K[y](t)
=
\int_0^\infty a(\tau)z_\tau^{\,2}(t)\,d\tau
\geq 0
\label{eq:hereditary-dissipation}
\end{equation}
is the irreversible hereditary dissipation.

Identity \eqref{eq:memory-power-decomposition} shows that the mechanical
power associated with the hereditary stress consists of two distinct
parts: the rate of change of the stored memory energy and the
non-negative dissipative contribution
\eqref{eq:hereditary-dissipation}. Consequently, the term
$D_K[y](t)$ entering the thermal balance
\eqref{eq:thermal-equation} represents only the irreversible part of
the hereditary mechanical power and is therefore thermodynamically
consistent.

The above representation requires a non-negative relaxation spectrum.
Equivalently, the kernel can be represented as a positive mixture of
exponential relaxation modes. Introducing $\lambda=1/\tau$, this
representation may be written in the Laplace form
\begin{equation}
K(t)=\int_0^\infty e^{-\lambda t}\,d\mu(\lambda),
\qquad d\mu(\lambda)\geq0.
\label{eq:positive-spectral-representation}
\end{equation}
Representation \eqref{eq:positive-spectral-representation} identifies
the class of completely monotone kernels. In particular, such kernels
satisfy
\[
(-1)^n K^{(n)}(t)\geq0,
\qquad n=0,1,2,\ldots,
\]
whenever the corresponding derivatives exist.

It is important to distinguish the assumptions required at different
levels of the analysis. The local well-posedness of the coupled
Volterra problem does not require complete monotonicity; the basic
conditions stated in Section~1 are sufficient for that purpose.
The stronger spectral condition
\eqref{eq:positive-spectral-representation} is invoked when the
hereditary term is interpreted thermodynamically through
\eqref{eq:memory-power-decomposition} and when particular kernel
classes are considered in the asymptotic analysis.

Two qualitatively different memory mechanisms will be compared below.
A short-memory response is represented by the exponential kernel
\begin{equation}
K_{\mathrm{exp}}(t)=k_0e^{-t/\tau_m},
\qquad k_0>0,\quad \tau_m>0,
\label{eq:exponential-kernel}
\end{equation}
whereas a long-memory response is represented by the power-law kernel
\begin{equation}
K_{\mathrm{pow}}(t)
=
\frac{k_0}{(1+t/\tau_m)^p},
\qquad k_0>0,\quad \tau_m>0,\quad 0<p<1.
\label{eq:power-law-kernel}
\end{equation}
Both kernels \eqref{eq:exponential-kernel} and
\eqref{eq:power-law-kernel} are non-negative and decreasing. They are
also completely monotone and therefore admit a positive spectral
representation of the form
\eqref{eq:positive-spectral-representation}. However, their tails are
fundamentally different:
\[
K_{\mathrm{exp}}(t)\sim e^{-t/\tau_m},
\qquad
K_{\mathrm{pow}}(t)\sim t^{-p}
\quad\text{as }t\to\infty.
\]
This difference in the decay of the memory kernel will be shown in
Section~4 to generate qualitatively different long-time relaxation
regimes of the same coupled thermomechanical model.
	
\section{\large\bf 3. Local well-posedness and positivity of temperature}

We now establish the local well-posedness of the coupled problem
defined by \eqref{eq:nonlocal-mechanical} and
\eqref{eq:thermal-equation}, subject to the initial conditions
\eqref{eq:initial-conditions}. At this stage, complete monotonicity of
the kernel is not required.

For convenience, introduce the Volterra operator
\begin{equation}
(\mathcal{V}_K v)(t)
=
\int_0^t K(t-s)v(s)\,ds.
\label{eq:volterra-operator}
\end{equation}
Using the Volterra operator \eqref{eq:volterra-operator} and setting
$v(t)=\dot y(t)$, the mechanical equation
\eqref{eq:nonlocal-mechanical} can be written as
\[
\eta(T(t))v(t)
+
(\mathcal{V}_K v)(t)
=
E(T(t))[\varepsilon(t)-y(t)]
-
R(y(t),T(t)).
\]
The internal variable is then recovered from
\[
y(t)=y_0+\int_0^t v(s)\,ds,
\]
whereas the temperature is governed by
\eqref{eq:thermal-equation}.

For the local theory of the complete coupled system, we additionally
assume that the hereditary dissipation functional
$D_K[y](t)$ is causal and locally Lipschitz with respect to the
history of $y$ on bounded subsets. This is a technical condition for
local solvability and is independent of the stronger thermodynamic
spectral assumptions introduced in Section~2.

\medskip

\noindent\textbf{Theorem 1.}
Assume that $E(T)$ and $\eta(T)$ are continuous and locally Lipschitz
for $T>0$, with $E(T)>0$ and $\eta(T)>0$, and that $R(y,T)$ is
continuous and locally Lipschitz in its arguments. Let
$K\in L^1_{\mathrm{loc}}(\mathbb{R}_+)$,
$\varepsilon\in C^1([0,T_f])$, and
$q_{\mathrm{ext}}\in C([0,T_f])$. Suppose also that $D_K$ is causal
and locally Lipschitz with respect to the history of $y$ on bounded
sets. Then, for every initial state
\eqref{eq:initial-conditions}, there exists a time $t_*>0$ such that
the coupled problem \eqref{eq:nonlocal-mechanical},
\eqref{eq:thermal-equation} admits a unique local solution
\[
y,T\in C^1([0,t_*]).
\]

\medskip

\noindent\textit{Proof.}
Fix a bounded neighbourhood of $(y_0,T_0)$ contained in the physical
region $T>0$. Since $\eta$ is continuous and strictly positive, there
exists a constant $\eta_*>0$ such that
\[
\eta(T)\geq\eta_*
\]
throughout this neighbourhood.

For $v_1,v_2\in C([0,\delta])$, the Volterra operator satisfies
\[
\|\mathcal{V}_K v_1-\mathcal{V}_K v_2\|_\infty
\leq
\|K\|_{L^1(0,\delta)}
\|v_1-v_2\|_\infty.
\]
Since
\[
\|K\|_{L^1(0,\delta)}\longrightarrow0
\qquad\text{as}\qquad
\delta\longrightarrow0,
\]
and since $E$, $\eta$, $R$, and $D_K$ are locally Lipschitz on bounded
sets, the coupled integral formulation for $(v,y,T)$ defines a
contraction on a sufficiently small time interval. Banach's
fixed-point theorem therefore yields a unique local solution. The
relations $v=\dot y$ and \eqref{eq:thermal-equation} then imply
$y,T\in C^1([0,t_*])$. This proves the assertion.
\hfill$\square$

\medskip

The thermodynamic representation established in Section~2 provides
an additional property of the solution. In particular,
\eqref{eq:hereditary-dissipation} gives
$D_K[y](t)\geq0$ for kernels admitting the positive spectral
representation \eqref{eq:positive-spectral-representation}.

\medskip

\noindent\textbf{Proposition 1.}
Under the assumptions of Theorem~1, suppose additionally that the
memory kernel admits the positive spectral representation
\eqref{eq:positive-spectral-representation} and that
$q_{\mathrm{ext}}(t)\geq0$. If $T_0>0$ and
$T_{\mathrm{env}}>0$, then the temperature remains positive
throughout the interval of existence:
\begin{equation}
T(t)>0,\qquad 0\leq t\leq t_*.
\label{eq:temperature-positivity}
\end{equation}

\medskip

\noindent\textit{Proof.}
Using $\eta(T)\dot y^{\,2}\geq0$,
\eqref{eq:hereditary-dissipation}, and
$q_{\mathrm{ext}}(t)\geq0$, the thermal equation
\eqref{eq:thermal-equation} yields
\[
C\dot T(t)
\geq
-h[T(t)-T_{\mathrm{env}}].
\]
If $h>0$, multiplication by the integrating factor
$e^{ht/C}$ gives
\[
T(t)
\geq
T_{\mathrm{env}}
+
\bigl(T_0-T_{\mathrm{env}}\bigr)e^{-ht/C}.
\]
Since $T_0>0$ and $T_{\mathrm{env}}>0$, the right-hand side is
strictly positive for all $t\in[0,t_*]$.

If $h=0$, the same differential inequality reduces to
$\dot T(t)\geq0$, and therefore
\[
T(t)\geq T_0>0.
\]
Hence \eqref{eq:temperature-positivity} holds throughout the interval
of existence.
\hfill$\square$

\medskip

Theorem~1 therefore guarantees local existence and uniqueness under
the minimal Volterra assumptions together with the local regularity
required of the hereditary dissipation functional, whereas
Proposition~1 shows that the physically admissible region $T>0$ is
invariant for thermodynamically dissipative memory. These results
provide the mathematical basis for the subsequent analysis of the
long-time relaxation regimes.
\section{\large\bf 4. Influence of the memory kernel on relaxation asymptotics}

We now examine how the class of the relaxation kernel affects the
long-time behaviour of the coupled thermomechanical system. The
analysis is restricted to the linearized dynamics near a stationary
state. Accordingly, the results below describe the relaxation
mechanism close to equilibrium and are not intended as global decay
estimates for the full nonlinear problem.

Let $(y_*,T_*)$ be a stationary state corresponding to a constant
prescribed strain $\varepsilon_*$. Introduce the perturbations
\[
u(t)=y(t)-y_*,
\qquad
\theta(t)=T(t)-T_*.
\]
Linearization of the coupled system about $(y_*,T_*)$ yields a
Volterra system of the form
\begin{equation}
\begin{aligned}
 \eta_* \dot u(t)
 + a_*u(t)
 + b_*\theta(t)
 + \int_0^t K(t-s)\dot u(s)\,ds &=0,\\
 C\dot\theta(t)
 + h\theta(t)
 - c_*\dot u(t) &=0,
\end{aligned}
\label{eq:linearized-system}
\end{equation}
where $\eta_*>0$ and the coefficients $a_*$, $b_*$ and $c_*$ are
determined by the derivatives of the constitutive functions at the
equilibrium state. System \eqref{eq:linearized-system} retains the
two-way thermomechanical coupling of the original model while
isolating the effect of the memory kernel on the relaxation process.

For the exponential kernel \eqref{eq:exponential-kernel}, the memory
term can be represented by an additional internal variable satisfying
a first-order differential equation. Hence
\eqref{eq:linearized-system} is equivalent to a finite-dimensional
autonomous linear system. If the corresponding characteristic roots
have negative real parts, there exist constants $M>0$ and
$\omega>0$ such that
\begin{equation}
 |u(t)|+|\theta(t)|
 \leq
 M e^{-\omega t}
 \bigl(|u(0)|+|\theta(0)|\bigr),
 \qquad t\geq0.
\label{eq:exponential-relaxation}
\end{equation}
Thus, the short-memory kernel
\eqref{eq:exponential-kernel} produces exponential relaxation under
the standard spectral stability condition.

The situation is qualitatively different for the power-law kernel
\eqref{eq:power-law-kernel}. Let
\[
u_0=u(0),\qquad \theta_0=\theta(0).
\]
Taking the Laplace transform of \eqref{eq:linearized-system} gives
\[
\begin{aligned}
\bigl(\eta_*s+a_*+s\widehat K(s)\bigr)U(s)+b_*\Theta(s)
&=\bigl(\eta_*+\widehat K(s)\bigr)u_0,\\
-c_*sU(s)+(Cs+h)\Theta(s)
&=C\theta_0-c_*u_0.
\end{aligned}
\]
Hence the characteristic determinant is
\begin{equation}
 \Delta(s)
 =
 \bigl(\eta_*s+a_*+s\widehat K(s)\bigr)(Cs+h)
 + b_*c_*s,
\label{eq:characteristic-relation}
\end{equation}
and
\[
U(s)=
\frac{
(\eta_*+\widehat K(s))(Cs+h)u_0
-b_*(C\theta_0-c_*u_0)
}{\Delta(s)}.
\]
Thus, the low-frequency behaviour of \(\widehat K(s)\) enters not only
the characteristic determinant but also the transformed solution
explicitly.

For the power-law kernel \eqref{eq:power-law-kernel},
\[
 K_{\mathrm{pow}}(t)
 \sim
 k_0\tau_m^p t^{-p},
 \qquad 0<p<1,
\]
and therefore
\[
 \widehat K_{\mathrm{pow}}(s)
 \sim
 A_p s^{p-1},
 \qquad
 A_p=k_0\tau_m^p\Gamma(1-p),
 \qquad s\to0^+.
\]
Consequently, \(s\widehat K_{\mathrm{pow}}(s)\sim A_p s^p\), so the
transformed problem contains a branch point at \(s=0\).

Assume that \(a_*>0\), \(h>0\), the equilibrium is asymptotically
stable, and the transformed resolvent admits analytic continuation to
a slit neighbourhood of the origin, with no competing singularity on
the closed right half-plane. These standard conditions permit the
small-frequency branch contribution to be inverted by contour
deformation. Since
\[
\Delta(s)=a_*h+hA_p s^p+o(s^p),
\qquad s\to0,
\]
we obtain, for \(u_0\neq0\),
\[
U(s)
\sim
\frac{A_pu_0}{a_*}s^{p-1}.
\]
Using the second transformed equation, the leading non-analytic part of
\(\Theta(s)\) is
\[
\Theta_{\mathrm{br}}(s)
\sim
\frac{c_*A_pu_0}{a_*h}s^p.
\]
The branch-point inversion formulas then yield
\[
u(t)
\sim
\frac{k_0\tau_m^p}{a_*}u_0\,t^{-p},
\qquad
\theta(t)
\sim
-\frac{p\,c_*k_0\tau_m^p}{a_*h}u_0\,t^{-p-1},
\qquad t\to\infty.
\]
Thus the mechanical perturbation inherits the exponent \(p\) directly
from the long-memory kernel, whereas the thermal perturbation decays one
algebraic order faster in the present coupling structure. If \(u_0=0\),
the leading \(t^{-p}\) mechanical term vanishes and the decay is no
slower. In all cases covered by the above stability and regularity
assumptions,
\begin{equation}
 |u(t)|+|\theta(t)|
 =
 O(t^{-p}),
 \qquad t\to\infty.
\label{eq:algebraic-relaxation}
\end{equation}
The asymptotic estimate \eqref{eq:algebraic-relaxation} is therefore a
direct consequence of the branch contribution generated by the long
tail of the kernel \eqref{eq:power-law-kernel}.

The comparison of \eqref{eq:exponential-relaxation} and
\eqref{eq:algebraic-relaxation} reveals the central qualitative
distinction. With the thermomechanical coupling and the remaining
parameters kept unchanged, the exponential kernel produces
exponential relaxation, whereas the power-law kernel produces
algebraic relaxation. Thus, within the linearized regime near
equilibrium, the class of the Volterra memory kernel determines the
qualitative class of the long-time relaxation asymptotics.
\section{\large\bf 5. Numerical illustration and thermomechanical interpretation}

To illustrate the kernel-dependent relaxation regimes established in
Section~4, we consider a post-loading relaxation process for a
thermoviscoelastic polymer or composite element. After the prescribed
loading stage, the internal deformation gradually returns toward its
equilibrium state, while the rate of this process depends on the
character of the material memory.

For the numerical illustration, the linearized near-equilibrium model
is used. Time is nondimensionalized by a characteristic relaxation
time $\tau$,
\[
\hat t=\frac{t}{\tau}.
\]
The remaining coefficients are normalized with respect to the
characteristic mechanical scale. The same dimensionless
thermomechanical parameters and the same initial perturbation are used
for both memory mechanisms. Thus, the only factor changed between the
two simulations is the class of the relaxation kernel.

The short-memory case is represented by the exponential kernel
\eqref{eq:exponential-kernel}, whereas the long-memory case is
represented by the power-law kernel
\eqref{eq:power-law-kernel} with $p=0.5$. The numerical parameters are
chosen for illustrative purposes and are not interpreted as
experimentally identified characteristics of a particular material.

\begin{figure}[ht]
\centering
\includegraphics[width=0.78\textwidth]{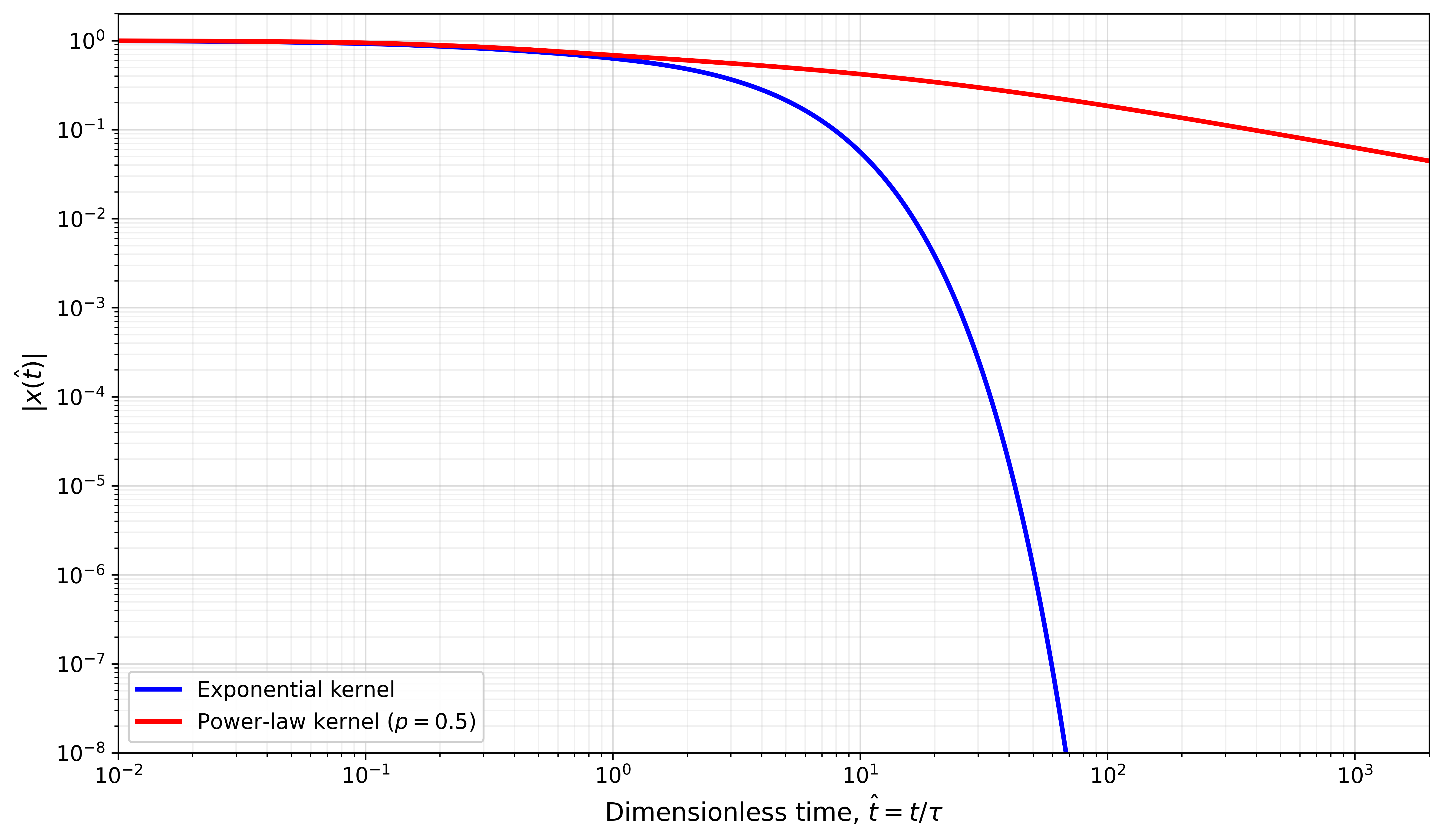}
\caption{Normalized mechanical relaxation for the exponential and
power-law memory kernels. The dimensionless time is
$\hat t=t/\tau$; for the power-law kernel, $p=0.5$.}
\label{fig:relaxation}
\end{figure}

Figure~\ref{fig:relaxation} shows the normalized mechanical relaxation
for the two kernels as a function of the dimensionless time $\hat t$.
A logarithmic scale is used on both axes in order to distinguish their
long-time behaviour.

The numerical curves in Figure~\ref{fig:relaxation} exhibit two
qualitatively different relaxation regimes. For the exponential
kernel, the perturbation decreases rapidly and becomes negligible
after a finite number of characteristic relaxation times. This
behaviour is consistent with the exponential estimate
\eqref{eq:exponential-relaxation}.

For the power-law kernel, the relaxation is substantially slower and
the response retains a visible long-time tail over the entire
computational interval. In accordance with
\eqref{eq:algebraic-relaxation}, the long-time part of the curve is
approximately linear on logarithmic axes, which is characteristic of
algebraic decay. For $p=0.5$, the theoretical asymptotic behaviour is
\[
|u(t)|\sim Ct^{-1/2},
\qquad t\to\infty.
\]

To quantify the algebraic decay, the numerical exponent is estimated
from two points $t_1$ and $t_2$ in the asymptotic interval according to
\begin{equation}
p_{\mathrm{num}}
=
-\frac{\ln |u(t_2)/u(t_1)|}
       {\ln(t_2/t_1)}.
\label{eq:numerical-decay-exponent}
\end{equation}
Using the interval
$1000\leq\hat t\leq2000$ gives
\[
p_{\mathrm{num}}\approx0.494,
\]
which is close to the theoretical value $p=0.5$. Thus, the numerical
experiment supports the algebraic relaxation law predicted by
\eqref{eq:algebraic-relaxation}.

For comparison, the exponential case yields a numerical slow
relaxation rate
\[
\lambda_{\mathrm{num}}\approx0.2679,
\]
in agreement with the theoretical value
$\lambda\approx0.26795$ for the chosen dimensionless parameters.
Hence the numerical calculation reproduces both characteristic
regimes: exponential decay for the short-memory kernel and algebraic
decay for the long-memory kernel.

From the thermomechanical point of view, the difference between the
two curves has a direct interpretation. An exponential kernel
represents a material with a dominant finite relaxation scale, so the
influence of earlier states is rapidly lost. A power-law kernel
represents a broad spectrum of relaxation times and preserves the
effect of remote history for considerably longer times. Consequently,
two material elements having the same instantaneous thermomechanical
parameters may display markedly different post-loading relaxation
solely because their memory spectra are different.

The numerical results therefore illustrate, rather than replace, the
analytical conclusion: keeping the thermomechanical coupling fixed
while changing the class of the memory kernel is sufficient to change
the qualitative long-time relaxation regime from exponential to
algebraic.
\section{\large\bf 6. Conclusions}

A coupled thermomechanical model with Volterra memory has been
formulated in which the internal mechanical variable and temperature
interact through temperature-dependent material properties and
mechanical dissipation. The hereditary contribution has been given
a thermodynamic interpretation by separating the energy stored in
the relaxation spectrum from the irreversible memory dissipation.
This representation also clarifies the role of completely monotone
kernels without imposing this stronger condition on the basic local
well-posedness problem.

Local existence and uniqueness of the coupled problem have been
established under minimal assumptions on the constitutive functions
and the Volterra kernel. For thermodynamically dissipative memory
kernels and a non-negative external heat source, positivity of the
absolute temperature is preserved throughout the interval of
existence.

The main result concerns the influence of the memory-kernel class on
the long-time dynamics near equilibrium. For the same linearized
thermomechanical coupling, an exponential kernel leads to exponential
relaxation, whereas a power-law kernel with a long tail produces
algebraic relaxation. Thus, changing the memory kernel while keeping
the remaining structure of the model fixed is sufficient to change
the qualitative class of the relaxation asymptotics.

The numerical illustration confirms this distinction for a common
set of dimensionless thermomechanical parameters. The exponential
kernel produces rapid fading of the perturbation, while the power-law
kernel retains a pronounced long-time tail consistent with the
predicted algebraic decay.

The obtained results are deliberately restricted to the asymptotic
behaviour of the linearized system near equilibrium and do not imply
global decay estimates for the full nonlinear problem. Extension of
the kernel-dependent asymptotic analysis to the nonlinear regime
constitutes a natural direction for further investigation.
	
	\vspace{2mm}
\noindent\textbf{Conflict of interest and ethics.} The author declares no conflict of interests. The author also declares full adherence to all journal research ethics policies, namely involving the participation of human subjects anonymity and/or consent to publish.

\vspace{2mm}
\noindent\textbf{Acknowledgements.} The author declares no special funding for this work.

\selectlanguage{english}
\bibliographystyle{amsplain}
\renewcommand{\refname}

\selectlanguage{ukrainian}
\vspace{2mm}
UDC 517.968.7
\vspace{2mm}
\begin{center}
\textbf{\large Зв’язана термомеханічна модель з пам’яттю типу Вольтерри: вплив класу ядра на асимптотику релаксації}
\end{center}
\vspace{1mm}
\begin{center}
\textbf{Андрій Сясєв}
\end{center}

\vspace{2mm}
\textit{Анотація.} Метою дослідження є встановлення впливу класу та асимптотичної поведінки ядра пам’яті типу Вольтерри на релаксаційні процеси у зв’язаній термомеханічній моделі. Модель описує взаємний вплив внутрішньої механічної змінної та температури через температурно-залежні характеристики матеріалу й механічну дисипацію. Дослідження поєднує інтегральну постановку Вольтерри, спектральне представлення механізму пам’яті, термодинамічний розклад спадкової потужності, аналіз локальної коректності, лінеаризацію поблизу стану рівноваги та асимптотичний аналіз із використанням перетворення Лапласа. Спектральне представлення дає змогу відокремити енергію, накопичену в релаксаційних модах, від необоротної спадкової дисипації та встановити термодинамічну узгодженість внеску пам’яті. За мінімальних припущень щодо визначальних функцій і ядра встановлено локальне існування та єдиність розв’язків; для термодинамічно дисипативної пам’яті та невід’ємного зовнішнього джерела тепла доведено збереження додатності абсолютної температури. Основний результат стосується лінеаризованої динаміки поблизу стану рівноваги: за незмінного термомеханічного зв’язку експоненціальне ядро зумовлює експоненціальну релаксацію, тоді як степеневе ядро з довгим хвостом приводить до алгебраїчної релаксації. Чисельне порівняння, виконане для однакового набору безрозмірних термомеханічних параметрів, ілюструє ці два асимптотичні режими. Отримані результати показують, що зміна лише класу ядра пам’яті може змінювати якісний закон довготривалої релаксації зв’язаної системи.

{\it Ключові слова:} зв’язана термомеханічна модель; пам’ять типу Вольтерри; ядро релаксації; спадкові ефекти; повна монотонність; експоненціальна релаксація; степенева пам’ять; асимптотика релаксації.

\renewcommand{\refname}{\bf Список використаних джерел}
\bibliographystyle{amsplain}

\vspace{4mm}
\selectlanguage{ukrainian}
\begin{center}
\textbf{Про автора / About the author}
\end{center}

\textbf{Андрій Сясєв}, кандидат фізико-математичних наук, доцент, кафедра математичного аналізу та оптимізації, Дніпровський національний університет імені Олеся Гончара, Дніпро, Україна;

\textbf{Andrii Siasiev}, Candidate of Physical and Mathematical Sciences, Associate Professor, Department of Mathematical Analysis and Optimization, Oles Honchar Dnipro National University, Dnipro, Ukraine.

\end{document}